\documentclass[10pt,aps,pre,twocolumn,showpacs,superscriptaddress]{revtex4-1}

\usepackage{amsmath}
\usepackage{amssymb}
\usepackage{amscd}
\usepackage{wasysym}
\usepackage[ansinew]{inputenc}
\usepackage[T1]{fontenc}
\usepackage{ae,aecompl}
\usepackage{sidecap}
\usepackage{hyperref}
   
\usepackage[pdftex]{graphicx}

\usepackage{color}
\definecolor{red}{rgb}{1,0,0}
\definecolor{lred}{rgb}{1,0.7,0.7}
\definecolor{lgreen}{rgb}{0.7,1,0.7}

\usepackage{colortbl}

\newcommand{\be}{\begin{equation}}
\newcommand{\ee}{\end{equation}}
\newcommand{\bea}{\begin{eqnarray}}
\newcommand{\eea}{\end{eqnarray}}

\newcommand{\ov}{\text{ov}}

\begin{document}
\bibliographystyle{apsrev}

\title{Nonlocal effects and counter measures in cascading failures}

\author{Dirk Witthaut}
\affiliation{Forschungszentrum J\"ulich, Institute for Energy and Climate Research -
	Systems Analysis and Technology Evaluation (IEK-STE),  52428 J\"ulich, Germany}
\affiliation{Institute for Theoretical Physics, University of Cologne, 
		50937 K\"oln, Germany}

\author{Marc Timme}
\affiliation{Network Dynamics, Max Planck Institute for Dynamics and Self-Organization (MPIDS), Am Fassberg 17, 37077 G\"ottingen, Germany}
\affiliation{Faculty of Physics, Georg August University G\"ottingen,
  Germany}

\date{\today}


\begin{abstract}
We study the propagation of cascading failures in complex supply 
networks with a focus on nonlocal effects occurring far away 
from the initial failure. 
It is shown that a high clustering and a small average path length of a network
generally suppress nonlocal overloads. These properties are typical for 
many real-world networks, often called small-world networks, such that 
cascades propagate mostly locally in these networks.
Furthermore, we analyze the spatial aspects of countermeasures based 
on the intentional removal of additional edges. Nonlocal actions are 
generally required in networks which have a low redundancy and are 
thus especially vulnerable to cascades.
\end{abstract}

\maketitle


\section{Introduction}

A reliable supply of electric power fundamentally underlies the 
function of most of our technical infrastructure and affects all 
aspects of daily life. Large-scale power outages can thus have 
potentially catastrophic consequences and cause huge 
economic losses \cite{Fair04,Amin05a}. 
Therefore it is an important goal to understand 
the vulnerability of a grid on all scales in order to secure our
energy supply. A promising direction is to combine methods 
and models of power engineering with the recent progress 
in the theory of complex networks \cite{Hill06,Newm11,Bara12,Brum13}.

Notably, most large-scale outages can be traced back to the 
failure of a single transmission element of our power supply 
system \cite{Pour06}. The initial failure then causes secondary 
failures in other elements of the grid and eventually a 
global cascade.
Cascading failures have been analyzed in a variety of studies
from the viewpoint of mathematics and theoretical physics
in the last decade  
\cite{Watt02,Mott02,Buld10,Albe04,Zhao04,Cruc04b,Heid08,Simo08,Schn11,Mott04,Scha06,Huan08}. 
It has been analyzed which structural properties of networks
promote or prevent global cascades 
\cite{Mott02,Watt02,Albe04,Zhao04,Cruc04b,Buld10}
and how fluctuations and transient dynamics affect the vulnerability 
of the grid \cite{Heid08,Simo08}.
Different countermeasures were discussed in order to make
a grid more robust beforehand \cite{Schn11} or to stop a
cascade before it affects major parts of the grid 
\cite{Mott04,Scha06,Huan08}. 

Most of these studies adopt a global perspective on
cascading failures and focus on the statistical properties of
the cascade and potential countermeasures. In this article
we study cascades from a more microscopic perspective and 
analyze the location and propagation of failures. In particular,
we characterize the nonlocality of secondary failures and 
show which structural features determine the nonlocality 
during the propagation of a cascade. It is shown that
overloads occur mostly locally, i.e. in the immediate 
neighborhood of the failing element, when the network is 
strongly clustered and `small'. Remarkably, these two
features are found for many real-world networks in
technology as well as in biology and sociology \cite{Watt98}.
We then extend these ideas to analyze the mechanism and
the spatial aspects of countermeasures based on the
intentional shutdown of transmission elements 
\cite{Mott04,Huan08}.

\section{Models for cascading failures}

To analyze the spatial aspects of cascading failures in complex
networks we use a model introduced by Motter and Lai in
\cite{Mott02,Mott04}. Related models were introduced and 
discussed in \cite{Holm02a,Holm02b,Cruc04}.
The Motter-Lai model assumes that at each time step, 
one unit of energy or information is sent from each vertex to each 
other vertex in the connected component along the shortest 
path. The load of each edge $F_{ij}$ is then given by the number 
of shortest paths running over this edge $i \leftrightarrow j$, 
which is nothing than the \emph{edge betweenness centrality} 
\cite{Newm10}. Furthermore, it is assumed that the capacity 
of each edge is proportional to the load of the edge in the initial 
intact network,
\be
   K_{ij} = (1+\alpha) F^{(0)}_{ij} \, ,
\ee
where the superscript $(0)$ denotes the initial intact network.
The tolerance parameter $\alpha \ge 0$ quantifies the global redundancy 
of the network: Each edge can transmit $(1+\alpha)$ of its initial load 
before it becomes overloaded. 

\begin{figure*}[tb]
\centering
\includegraphics[width=16cm, angle=0]{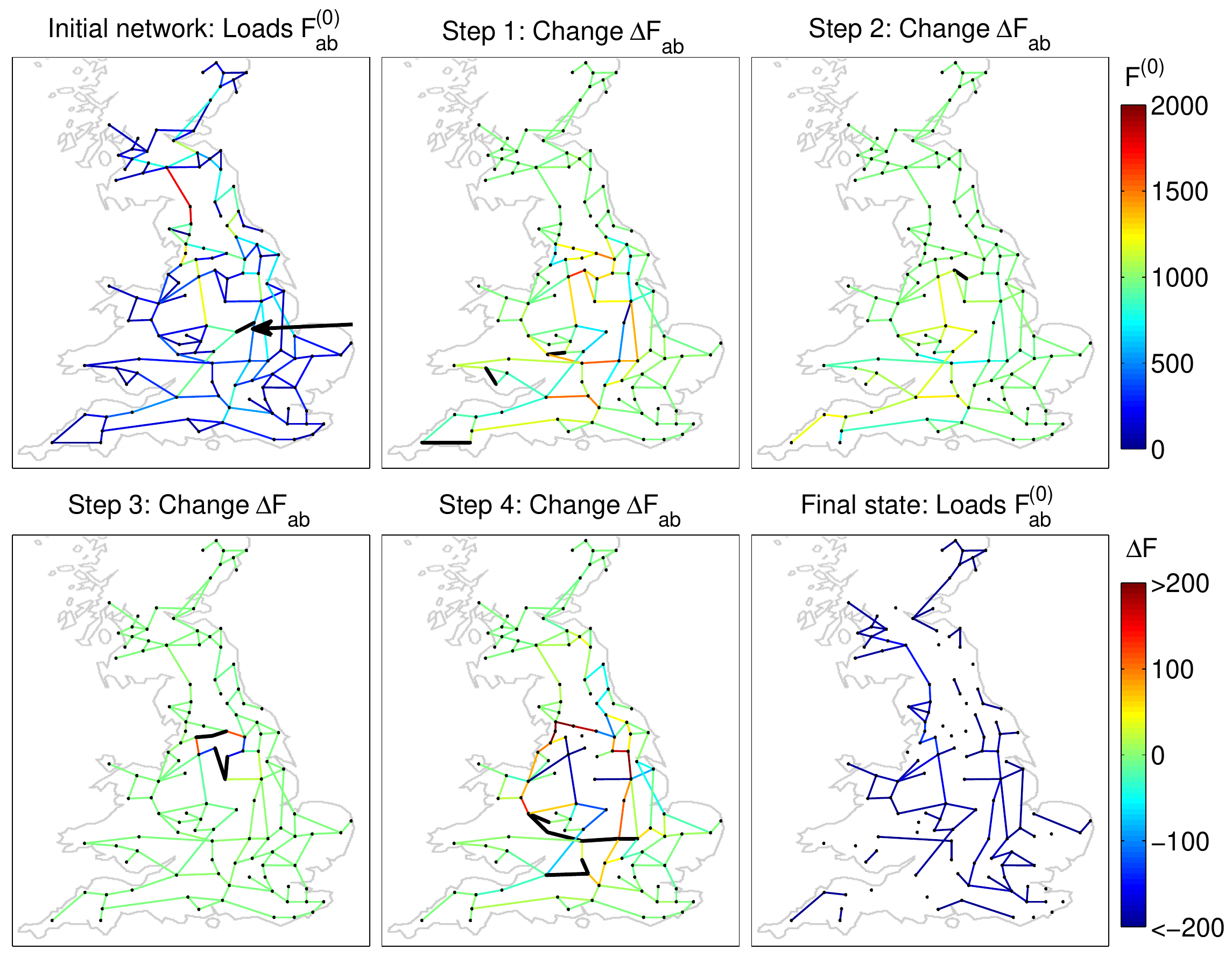}
\caption{
\label{fig:cascade1}
(Color online)
Propagation of a cascade of failures in the British high-voltage 
power transmission grid. The cascade is triggered by a single edge
which drops out of operation (marked by an arrow).
As a consequence, the network flow is rerouted which causes
overloads of further edges which also break down (thick black lines).
Most interestingly, the overloads do not occur in the immediate
neighborhood of the removed edge. The network structure was
taken from \cite{Simo08,12powergrid} and the tolerance parameter 
is $\alpha = 0.5$.
}
\end{figure*}      

Then it is analyzed what happens if one edge is damaged, 
such that it is effectively removed from the network. 
Obviously, the other edges have to take over the load such 
that $F_{ij}$ will generally increase. If the load exceeds the 
capacity of an edge $(i,j)$, $F_{ij} > K_{ij}$, then this edge becomes
overloaded and also drops 
out of service, which causes a further redistribution of the flows
and further overloads. This can trigger a large cascade 
of failures disconnecting the entire grid.
We note that the original articles  \cite{Mott02,Mott04} analyze 
potential overloads of vertices instead of edges. However,
in cascading failures of power grids, usually the transmission
lines (i.e. the edges) become overloaded and drop out of service. 
Therefore we concentrate on edges instead of vertices in
the present paper.

An example of a cascading failure in the Motter-Lai model is shown
in Fig.~\ref{fig:cascade1} for the topology of the British 
high-voltage transmission grid \cite{Simo08,12powergrid}.
The cascade is triggered by the breakdown of one edge 
marked by an arrow in the upper left panel of the figure.
The cascade then propagates through the network and
finally leads to a state where the network is decomposed
into several components.

A remarkable aspect of this example is that the cascade
is strongly \emph{nonlocal}. The distance of the defective
edge causing the flow redistribution and the overloaded
edges is rather large. Therefore a local perspective is not
sufficient to evaluate the effects of the breakdown of 
single edges in a complex network. In the following we
will analyze the spatial aspects of cascading failures in detail
and show which topologies are especially prone to nonlocal 
failures.

On a \emph{global} scale, the damage caused by a cascading 
failure is generally quantified by the number of vertices which 
are still connected when the cascade comes to a halt. To be 
precise, we measure the number of vertices in the largest 
connected component in the final state (called $G$) as well
as in the initial network (called $G_0$). A high value of the
ratio $G/G_0$ indicates that the network is still mostly 
intact, while a low value value of $G/G_0$ indicates a
fatal global cascade.
Numerical results for the average effect of cascading failures 
are shown in Fig.~\ref{fig:alpha} (a,b) as a function of the tolerance 
parameter $\alpha$. Obviously the size of the final cluster
$G/G_0$ increases with $\alpha$ -- in general,  catastrophic 
global cascades are more likely in networks that lack redundancy,
i.e. for low values of $\alpha$.
This plot also shows which amount of redundancy is needed in 
order to contain the possible effects
to a maximum acceptable value. How the network topology
determines these curves and thus the global robustness of
a network has been discussed intensively in the literature
(see, e.g., \cite{Mott02,Watt02,Albe04,Zhao04,Cruc04b,Buld10}).
However, such an analysis does not reveal which parts of the 
network are prone to outage and how a cascade propagates
on a microscopic level.

\section{Nonlocality of cascading failures}

\begin{figure}[tb]
\centering
\includegraphics[width=8cm, angle=0]{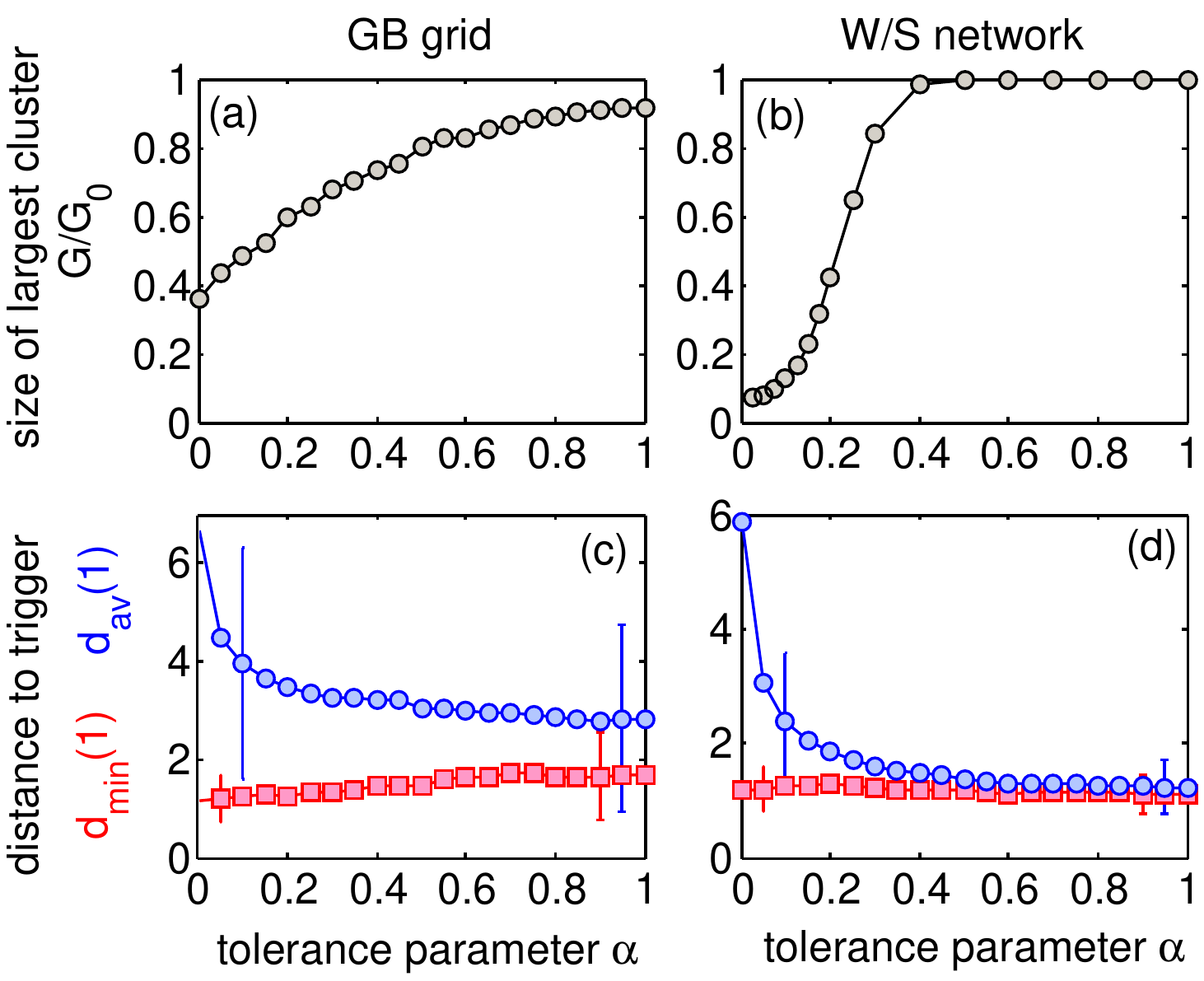}
\caption{\label{fig:alpha}
(Color online)
Resilience and nonlocality in cascading failures as a function of
the tolerance parameter $\alpha$.
(a,b) Relative size of the largest connected cluster after the
cascade $G/G_0$, averaging over all possible trigger edges.
(c,d) Average of the distance to the trigger edge for all 
overloaded edges ($d_{\rm av}(1)$, $\circ$) and for the edge which is nearest 
to the trigger ($d_{\rm min}(1)$, $\square$) for the first step of the cascade.
The vertical bars show typical values
for the respective standard deviation.
Results are collected for all possible trigger edges in the respective 
network:
(a,c) the British power grid \cite{Simo08,12powergrid} and 
(b,d) a W/S network with $N = 500$ vertices, $k=4$ and
$q = 0.2$ \cite{Watt98}.
}
\end{figure} 

\begin{figure}[tb]
\centering
\includegraphics[width=8cm, angle=0]{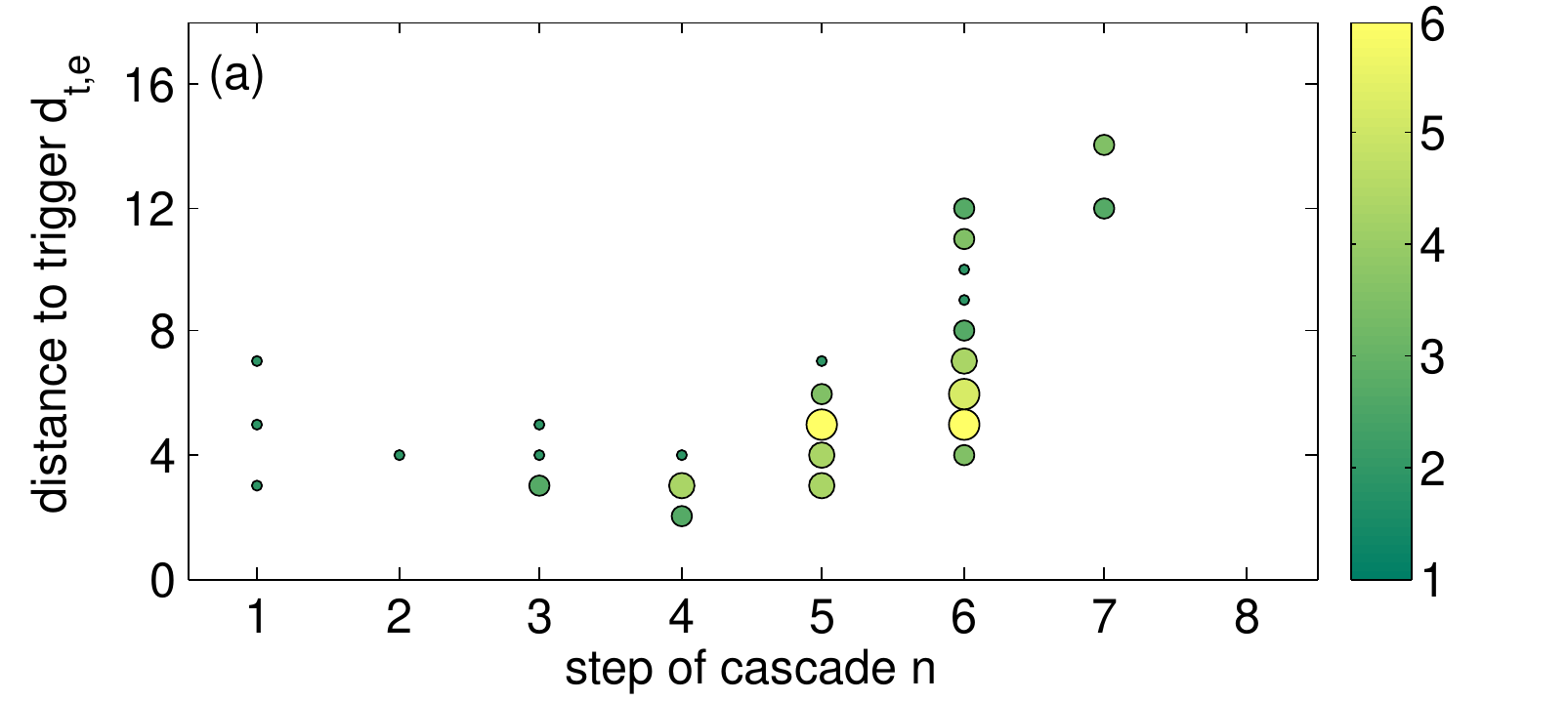}
\includegraphics[width=8cm, angle=0]{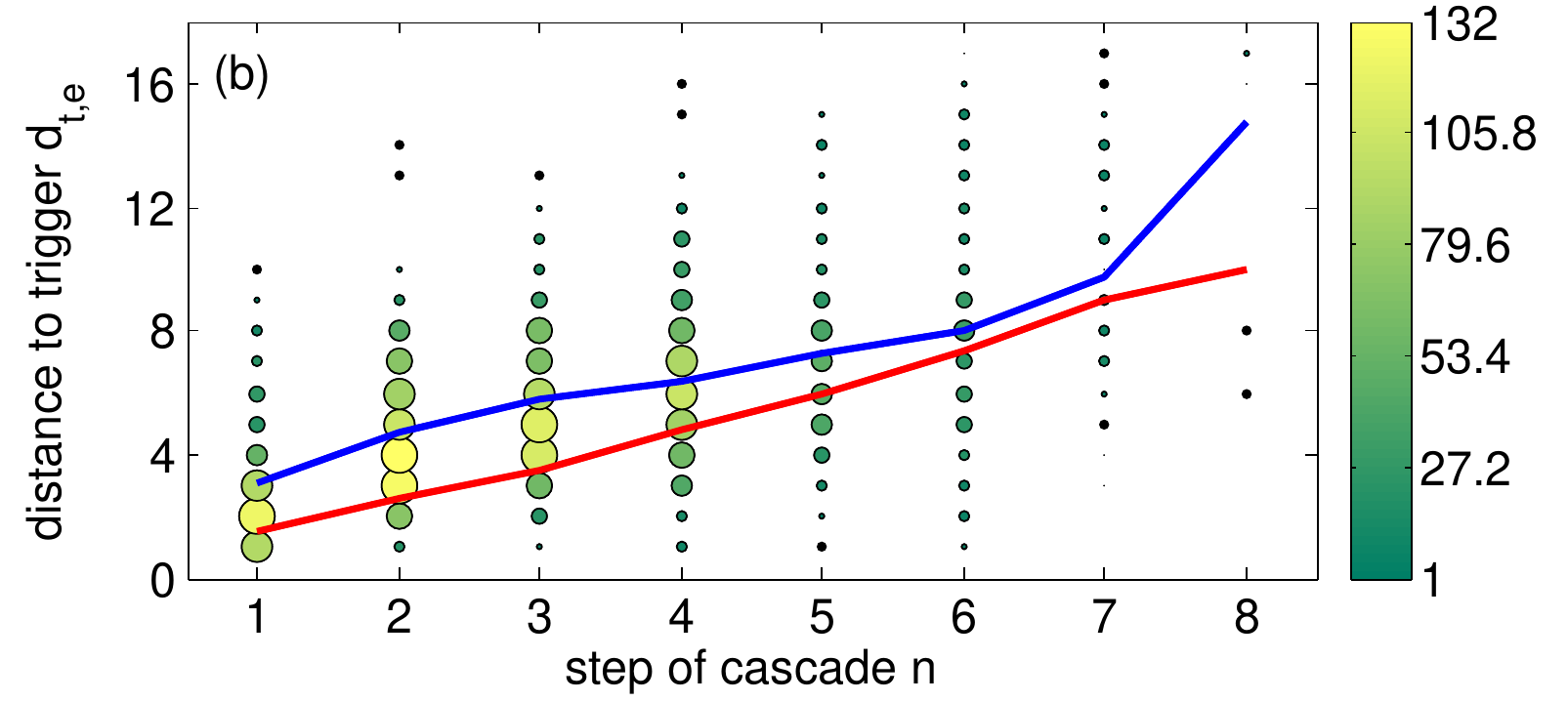}
\caption{\label{fig:dvsn}
(Color online)
Propagation of a cascade of failures in the British high-voltage 
power transmission grid. The color color and the area of the symbols
indicates the number of edges that are overloaded at step $n$ and 
located at a 
distance $d_{t,e}$ from the trigger edge.
(a) Data for a single cascade as shown in Fig.~\ref{fig:cascade1}.
(b) Data collected for all possible trigger edges in the same network.
The blue line is the average distance to the trigger
$d_{\rm av}(n)$ and the red line is distance of the 
\emph{nearest} overloaded edge $d_{\rm min}(n)$. 
}
\end{figure} 

To analyze the nonlocality of failures in complex networks
we must first specify the meaning of `distance' in a network.
The distance $d_{a,b}$ of two vertices $a$ and $b$ is defined as
as the number of edges in a shortest path connecting them 
\cite{Jung12}. 
Furthermore we need the distance of two \emph{edges}
$(a,b)$ and $(c,d)$, which is defined as the number of 
vertices on a shortest path between the edges such that
\be
   d_{(a,b),(c,d)} = \min_{x\in \{a,b\}, y\in \{c,d\}} d_{x,y} +1.
  \label{eqn:def-edgedist}
\ee
In the following we denote by $t$ the edge whose initial break-
down triggers the cascade and by $\ov(n)$ the set of all edges
overloaded at the nth step of the cascade.
We then analyze the distribution of the distances $d_{t,e}$ for
all overloaded edges $e \in \ov(n)$ as well as its average
\be
   d_{\rm av}(n) = \Big\langle \big\langle d_{t,e} \big\rangle_{e \in \ov(n)} \Big\rangle_{t} \, .
\ee
Furthermore, we analyze where the nearest overload occurs during 
the $n$th step, i.e the minimum of the distance between the 
trigger $t$ and all edges $e \in \ov(n)$.
This quantity is calculated separately for each cascade and we take
the average over all potential trigger edges, 
\be
   d_{\rm min}(n) = \Big\langle \min_{e \in \ov(n)}  d_{t,e}  \Big\rangle_t \, .
\ee

The distance between overloaded edges and the initial trigger 
edge is shown in Fig.~\ref{fig:dvsn} (a) for the example shown 
in Fig.~\ref{fig:cascade1}.
Already in the first step $n=1$ we observe three overloaded edges
at distances $d = 3,5,7$, i.e. at rather remote locations.
In the following we will concentrate on this first step of the
cascade which facilitates the understanding of nonlocal effects.
In later steps $n>1$ of the cascade there are generally multiple failures 
occurring at once. Further outages then occur due to the collective 
redistribution of network flows and cannot be attributed to a 
single cause alone.
Quantifying the \emph{direct} nonlocality of flow rerouting, i.e the nonlocality 
from one step of a cascade to the next step, thus faces conceptual 
difficulties except for step $n=1$.
The distance of overloaded edges to the initial trigger edge shown
in Fig.~\ref{fig:dvsn} accounts for the \emph{indirect} 
nonlocality of a cascade for $n>1$, as it includes the propagation 
over several intermediate steps.

The influence of the global redundancy of a network on the nonlocality 
of flow rerouting is analyzed in Figure \ref{fig:alpha} (c,d). We plot the
distance between the overloaded edges and the  trigger 
edge $d_{\rm av}(n)$ and $d_{\rm min}(n)$ for $n=1$ (the direct nonlocality) as a 
function of the tolerance parameter $\alpha$. 
The first quantity shows
where typical overloads occurs, while the latter quantity shows 
where the \emph{nearest} overload occurs. 
It is observed that the average distance between overload and 
trigger $d_{\rm av}(1)$ decreases strongly as a function of the tolerance parameter $\alpha$.
In highly redundant networks, i.e. for large values of $\alpha$, a large change of the flow 
$F_{i,j}$ is needed to induce an overload. Such changes are 
rare and occur almost exclusively in the neighborhood of the trigger. 
The average distance between trigger and overload $d_{\rm av}(1)$ 
is small and the rare cascades propagate 'locally'. 
In weakly redundant networks, i.e.~for small value of $\alpha$, already medium 
scale changes of the flow $F_{i,j}$ induce overloads. Such changes occur 
frequently also in remote areas of the network. The average
distance to the trigger $d_{\rm av}(1)$ is large and cascades 
can be strongly nonlocal. Such events are hard to predict and to contain. 

\section{The role of network topology}

\begin{figure}[tb]
\centering
\includegraphics[width=8cm, angle=0]{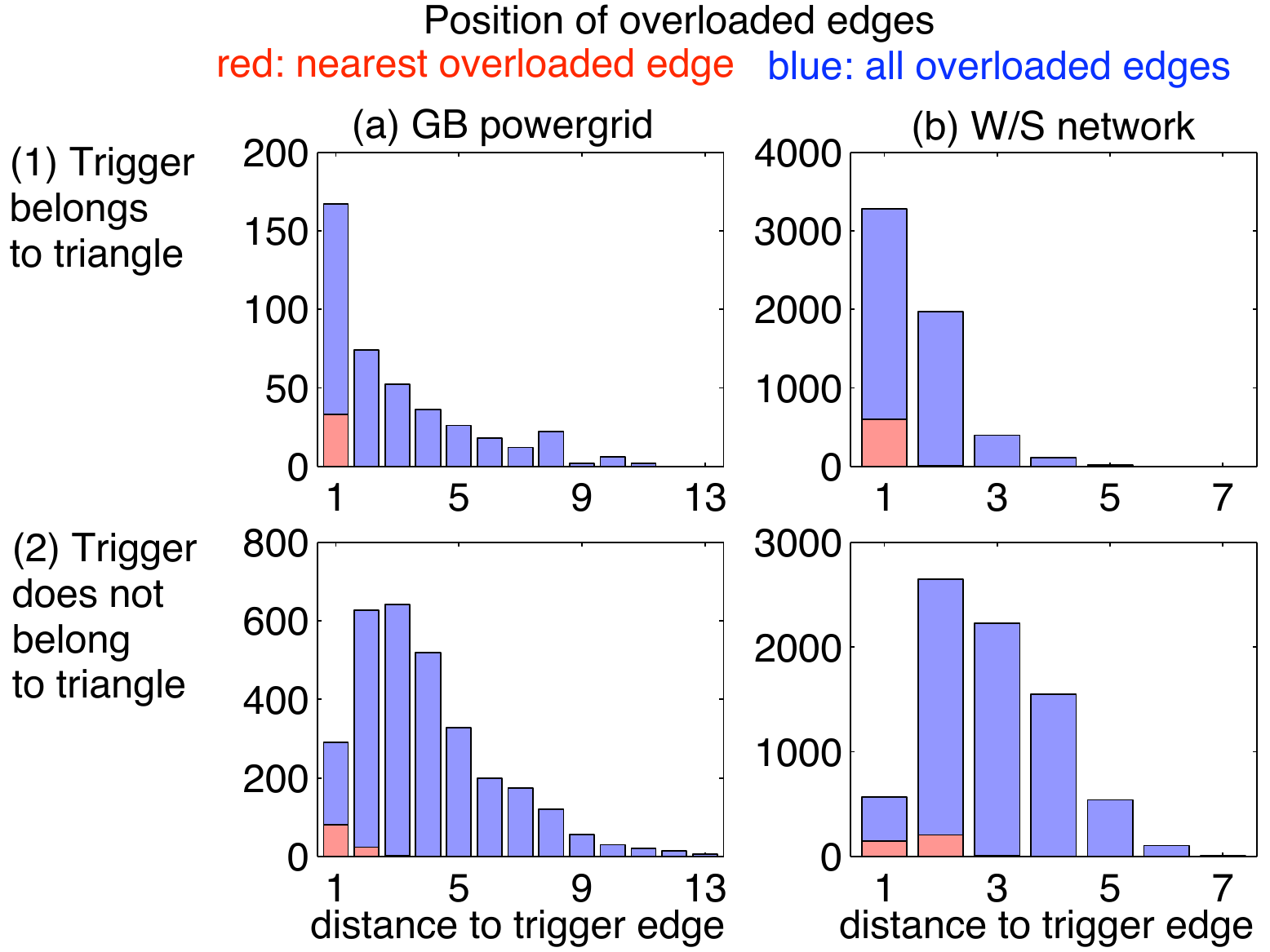}
\caption{\label{fig:motif}
(Color online)
The position of overloaded edges after the failure of a single
trigger edge strongly depends on whether the trigger belongs
to a triangle (upper panels) or not (lower panels).  
We plot a histogram of the distances to the trigger edge for all 
overloaded edges ($d_{t,e}$ for all $e \in \ov(1)$, blue) and for 
the edge which is nearest to the trigger 
($\min_{e \in \ov(1)} d_{t,e}$, red). 
Results are collected for all possible trigger edges in the respective 
network:
(a) the British power grid with $\alpha=0.1$ \cite{Simo08,12powergrid}, 
(c) a W/S network with $N = 500$ vertices, $k=4$, 
$q = 0.2$ and $\alpha=0.1$ \cite{Watt98}.
}
\end{figure} 

\begin{figure*}[tb]
\centering
\includegraphics[width=12cm, angle=0]{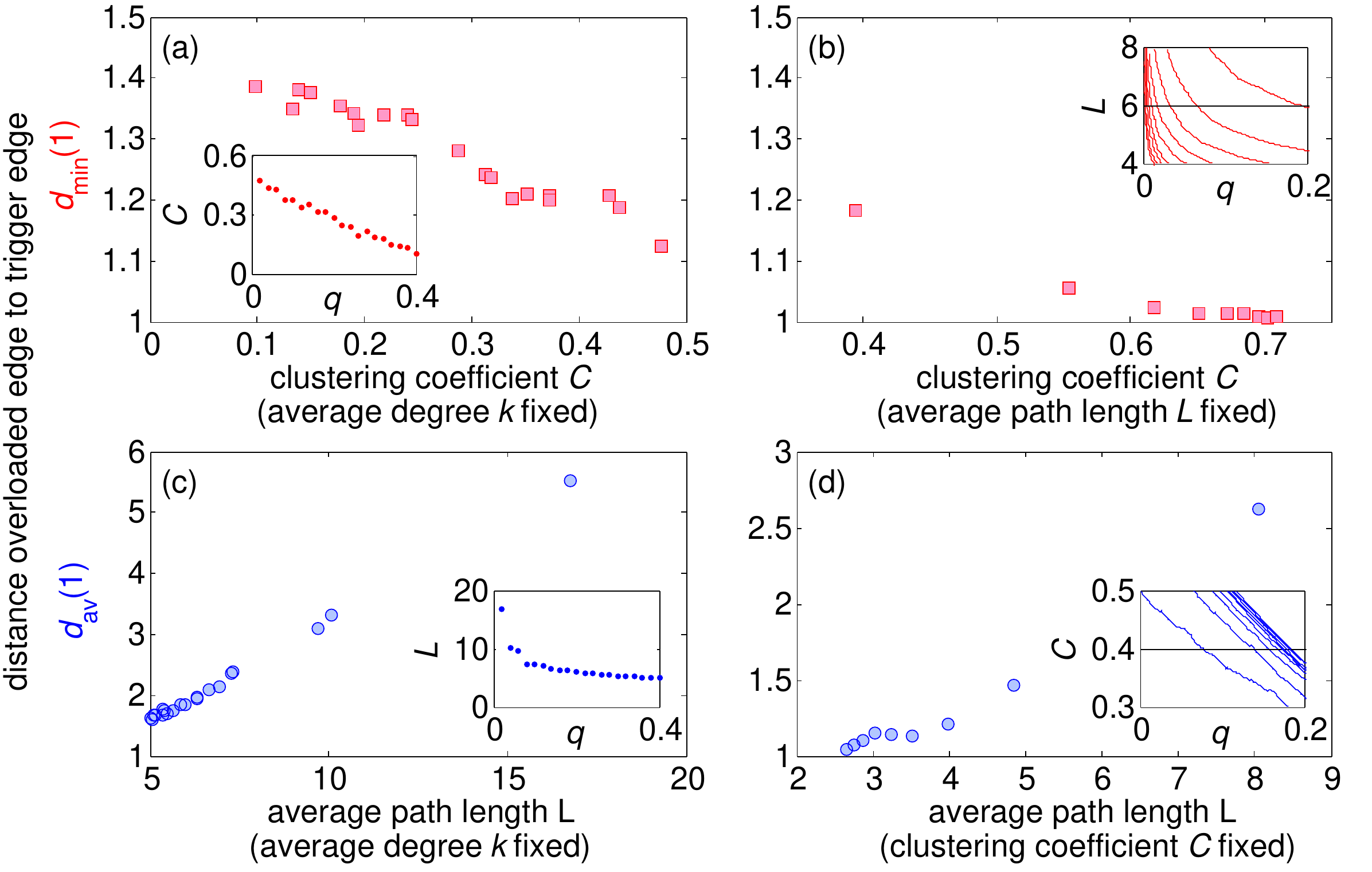}
\caption{\label{fig:cluster}
(Color online)
Nonlocality as a function of the network topology.
(a,b) The average distance of the nearest overloaded edge to the trigger 
$d_{\rm min}(1)$  decreases with the clustering coefficient $C$ of the network.
(c,d) The average distance of all overloaded edges $d_{\rm av}(1)$ increases with
the average path length $L$.
In panels (a,c) results are shown for W/S networks with fixed average 
degree $k=4$ and different values of the topological randomness $q$. 
The insets show how $C$ and $L$ scale with $q$.
In panels (b,d) results are shown for W/S networks with different degrees 
$k=4,6,\ldots,20$. The topological randomness $q$ has been chosen 
such that either the path length is fixed as $L \approx 6$ in (b) or
the clustering coefficient is fixed as $C \approx 0.4$ in (d), cf.~the
insets. The network has $N=500$ nodes and $\alpha = 0.2$.
}
\end{figure*} 

The network topology has a decisive influence on the collective 
dynamics of complex networks, in particular the spread of 
information or perturbations (see \cite{Stro01,Bocc06,Newm10} 
and references therein).
The nonlocality of cascades of failures is essentially determined
by two topological features of the grid:
(1) the \emph{size} of the network which is measured by the 
average shortest path length 
\be
    L := \langle d_{x,y} \rangle_{x,y}
\ee
where the average is taken over all pairs of nodes $x,y$ and
(2) the \emph{availability of short redundant pathways} 
in the network. Such short paths are especially available
if the trigger edge belongs to a \emph{triangle} \cite{footnote}. 
On a global scale the presence of triangles in the network is
quantified by the clustering coefficient 
\cite{Watt98} 
\be
    C := \frac{3 \times \mbox{number of triangles}}
   {\mbox{number of connected triplets of vertices}} \, .
   \label{eqn:def-cluster}
\ee
These conclusions hold for individual cascades in a given network 
(cf.~Fig.~\ref{fig:motif}) and for average cascades in networks with
variable topology (cf.~Fig.~\ref{fig:cluster}).

We first consider individual cascades for a given network topology in more detail.
When a single trigger edge $(a,b)$ breaks down, the flow
$F_{a,b}$ has to be rerouted via an alternative path in the
network. This may cause an overload and thus a secondary 
failure at another edge $(i,j)$. Such an overload can happen
locally, i.e. in the direct neighborhood  of the trigger edge
defined by $d_{(a,b),(i,j)} = 1$ but also at a remote location
in the network. 
The location of potential overloads is determined by the 
location of the alternative paths which take over the load.
In particular, a \emph{short alternative path} is available 
when the vertices $a$ and $b$ belong to a closed 
\emph{triangle} $(a,b,c)$ \cite{footnote}.
Then there is an alternative path of length $2$ given by 
$a - c - b$, which will take over most of the flow $F_{a,b}$
when the edge $(a,b)$ fails. In this case it is very likely that 
an overload occurs locally at the two edges $(a,c)$ and 
$(c,b)$.

A statistical analysis of individual cascades confirms this claim. 
Figure \ref{fig:motif} shows histogram of the 
distance between the overloaded edges and the trigger edge,
where we distinguish if the trigger belongs to a triangle or not. 
Results are shown for all overloads as well as for the nearest 
overload. 
If the trigger edge belongs to a triangle (upper panels), the nearest overload 
occurs almost always within the triangle, i.e. at a distance of one. 
Further overloads can occur at different positions, but the probability 
decreases strongly with the distance.
On the contrary, nonlocal overloads are much more frequent if
the trigger edge does not belong to a triangle (lower panels).
The highest number of overloads is found not in the immediate
neighborhood of the trigger edge but at distance of $d=2$ or $d=3$. 
In this case the redistribution of the flow $F_{ab}$ cannot be 
predicted within  a simple local picture. 

To analyze how global structural properties of a network
determine the nonlocality of cascading failures we simulate 
cascades for an ensemble of networks that interpolate between 
regular and random 
structures introduced by Watts and Strogatz \cite{Watt98}, 
which are referred to as W/S networks in the following. To generate 
such a network one starts  with a ring, where each of the $N$
vertices is connected to its $k$ neighbors, $k$ being the average 
degree of the network. The total number of edges in the network
is thus given by $Nk/2$. Then a fraction $q$ of all edges is
randomly selected, deleted and re-inserted at a random 
position in the network. 
To reveal the influence of the size $L$ and the clustering coefficient $C$ , 
we study two cases in detail:
(1) W/S networks with a fixed value of $k$ and different topological 
randomness $q$, which affects both $C$ and $L$ simultaneously 
(Fig.~\ref{fig:cluster} a, c) and 
(2) W/S networks where where either $C$ or $L$ is kept constant by 
varying $k$ and $q$ simultaneously 
(Fig.~\ref{fig:cluster} b, d). 

The position of the \emph{nearest} overload is essentially determined by 
the  clustering coefficient $C$ which measures the probability 
that the trigger edge belongs to a triangle.
Indeed, we observe a strong decrease of the distance $d_{\rm min}(1)$ with
increasing clustering coefficient $C$ (cf.~figure \ref{fig:cluster} (a,b)).
This holds regardless of the fact whether we keep the degree $k$
or the average path length $L$ fixed.

\begin{figure*}[tb]
\centering
\includegraphics[width=16cm, angle=0]{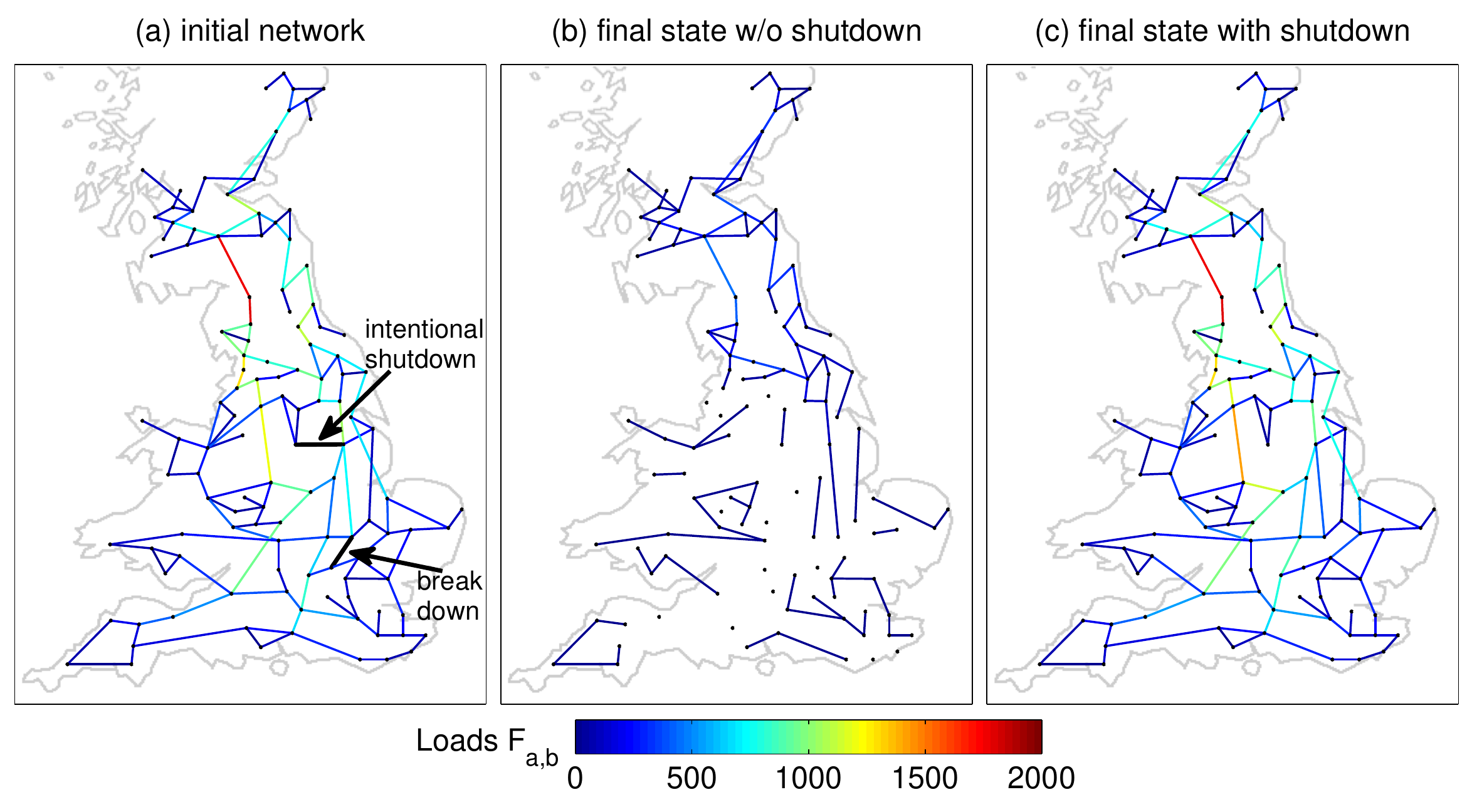}
\caption{\label{fig:prevent1}
(Color online)
Preventing a cascading failure by the intentional removal of 
a second edge within the Motter-Lai model.
(a) We consider a scenario where one edge breaks down. As a 
counter measure a second carefully selected edge is shut down.
The two edges are marked by arrows.
(b) Without any counter measure the initial breakdown
triggers a cascade of failures fragmenting the network.
(c) The cascade can be completely prevented if a second, carefully
chosen edge is also shut down (intentional removal, IR). When the 
two marked edges are removed simultaneously, no cascade takes place and the 
network remains fully connected.  The network structure was
taken from \cite{Simo08,12powergrid} and the tolerance parameter 
is $\alpha = 0.5$.
}
\end{figure*} 

The size of a network $L$ obviously limits the distances of vertices and
edges. The numerical results plotted in figure \ref{fig:cluster} (c,d)
reveal a much stronger influence. The \emph{average} distance of the 
overloaded edges to the trigger $d_{\rm av}(1)$ increases almost linearly with 
the average path length $L$. Only for very small values of $L$ 
does the distance saturate slightly below the lower limit $1$.
This result holds regardless of the fact whether we keep the 
degree $k$ or the clustering coefficient $C$ fixed.

We conclude that nonlocal overloads are particularly likely
if the network is weakly clustered and has a large average 
path length. Remarkably, many real-world networks from
power grids to biological and social network are so-called 
\emph{small worlds} in the sense that  both the clustering 
is high and the average path length is low. This small-world
regime is recovered in the W/S network ensemble for intermediate 
values of the topological randomness $q$ \cite{Watt98}. 
Our results suggest the conclusion that such small-world networks
are particularly \emph{local} in the sense that the probability
for  nonlocal failures is smallest. This result may provide an
additional reason why many real-world network have 
small-world properties (cf.~the discussion in \cite{Menc13}).

\section{Preventing cascades by intentional removal}

\begin{figure}[b]
\centering
\includegraphics[width=8cm, angle=0]{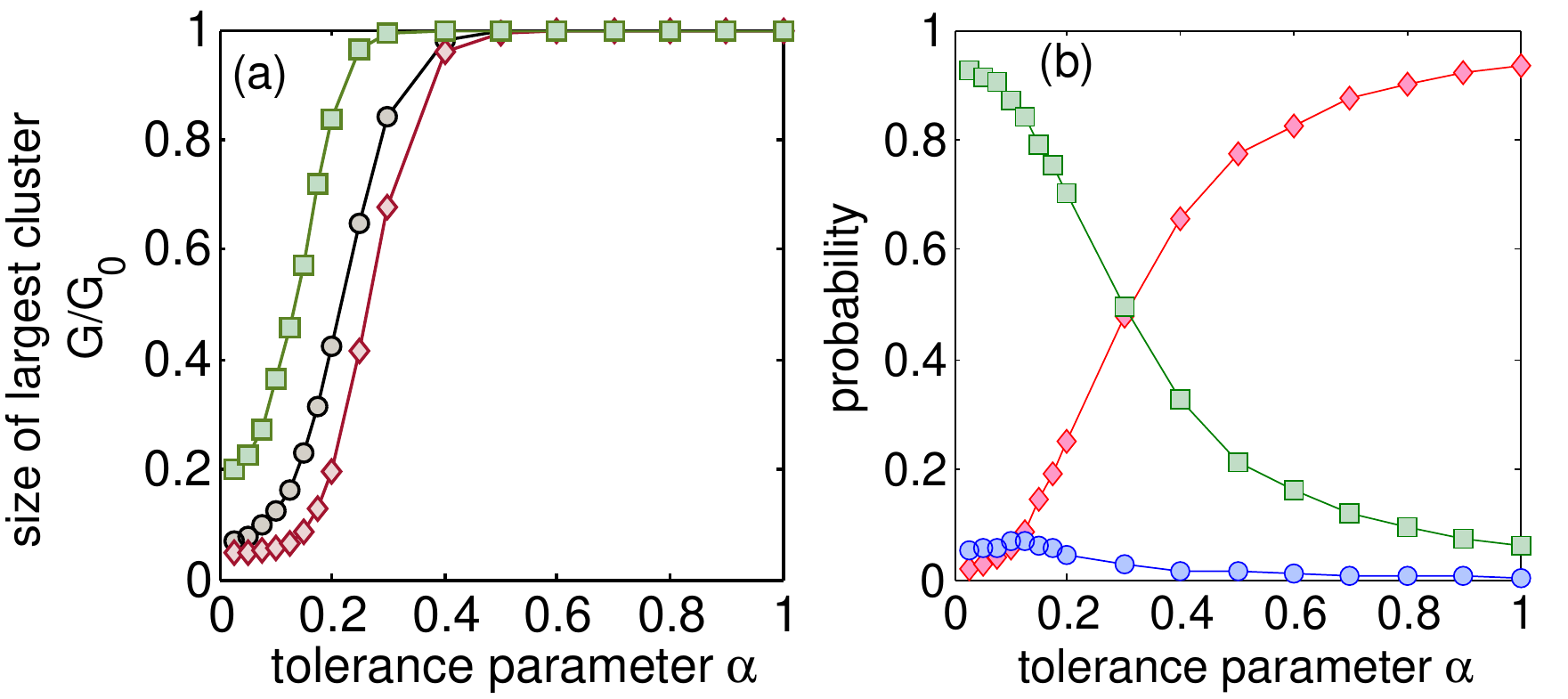}
\caption{\label{fig:ir1}
(Color online)
Effectivity of intentional removal (IR) for preventing cascading failures
as a function of the tolerance parameter $\alpha$.
(a) Relative size of the largest connected cluster after the
cascade $G/G_0$ averaging over all possible trigger edges.
We compare cascades triggered by $N-1$ errors ($\circ$) and
uncorrelated $N-2$ errors ($\diamond$) to the effect of an 
optimized intentional removal ($\square$).
(b) Probability that IR leads to an increase of the final cluster
size without disconnecting the network (green squares) 
in comparison to the probability 
that IR has no effect (red diamonds) and
that IR disconnects the grid (blue circles).
Results are collected for all possible trigger edges in a W/S 
network with $N = 500$ vertices, $k=4$ and $q = 0.2$ \cite{Watt98}.
}
\end{figure}

An effective counterstrategy for preventing global cascades 
of failures is the intentional removal (IR) of parts of the network
\cite{Mott04,Huan08}. 
Similar actions are taken in real-world power grids in case of
an emergency. If the power is no longer balanced in one part 
of the grid, for example after a cascade of transmission line 
failures, several consumers are actively disconnected 
(see, e.g., \cite{UCTE07}).
An example for a successful application of this strategy is shown
in figure \ref{fig:prevent1} where the removal of one additional edge
prevents the cascade completely. A statistical analysis of the
effectiveness of IR in the Motter/Lai-model is shown in 
Fig.~\ref{fig:ir1} for a W/S network.
We compare the effect of an optimized IR to cascades triggered
by the breakdown of a single edge (called $N-1$ errors) and
the uncorrelated simultaneous breakdown of two edges 
(called $N-2$ errors). Remarkably,
IR can reduce the number of disconnected vertices by more 
than $50 \%$ for intermediate values of the tolerance parameter $\alpha$. 

Two basic mechanisms contribute to the effectiveness of intentional 
removal. 
First, a small part of the network can be intentionally disconnected
by removing a single edge. This is possible if this part of the network 
is connected to the rest through a single edge only, which is then 
called a \emph{bridge} \cite{Jung12}.
In the Motter-Lai model each vertex transmits one unit of information 
or energy to all other vertices in the connected component. If several
vertices are disconnected they do no longer send or receive information
or energy from the rest such that the overall network flow decreases.
This method can be used to limit the consequences to a small
local outage instead of a global cascade. This can be very effective
in practice, but in any case parts of the network become 
disconnected.

However, in many cases there are much more sophisticated methods to 
prevent or stop a cascade of failures. An example is shown 
in Fig.~\ref{fig:prevent1}, where the breakdown of a single edge 
causes a cascade of failures leading to a strong fragmentation of 
the network. On the contrary, the intentional removal of another 
edge at a distance of $2$ prevents the cascade completely. 
In these cases the intentional removal of an edge leads to a collective redistribution
of the network flows which is beneficial and improves network stability.
We conclude that the removed edge is actually counterproductive as
it degrades network stability. This is analog to Braess' paradox, where
the addition of new edges in a supply or traffic network worsens
its operation or makes a network unstable
\cite{Brae68,Nish10,12braess,13nonlocal}.
Preventing cascades by intentional removal can thus 
be seen as an application of Braess' paradox.

The effectiveness of optimal IR is further analyzed in Fig.~\ref{fig:ir1} (b) 
as a function of the tolerance parameter $\alpha$ for a W/S network.
For a low value of the $\alpha$, IR is very effective in most cases
and does \emph{not} rely on the intentional disconnection of parts
of the grid. For the given network topology, this holds for more 
than $90 \%$ of all possible trigger edges.
For high values of $\alpha$, most initial failures do not lead
to a cascade at all. Consequently, IR has no effect with a very high 
probability -- simply because it is not needed. 

\begin{figure}[tb]
\centering
\includegraphics[width=8cm, angle=0]{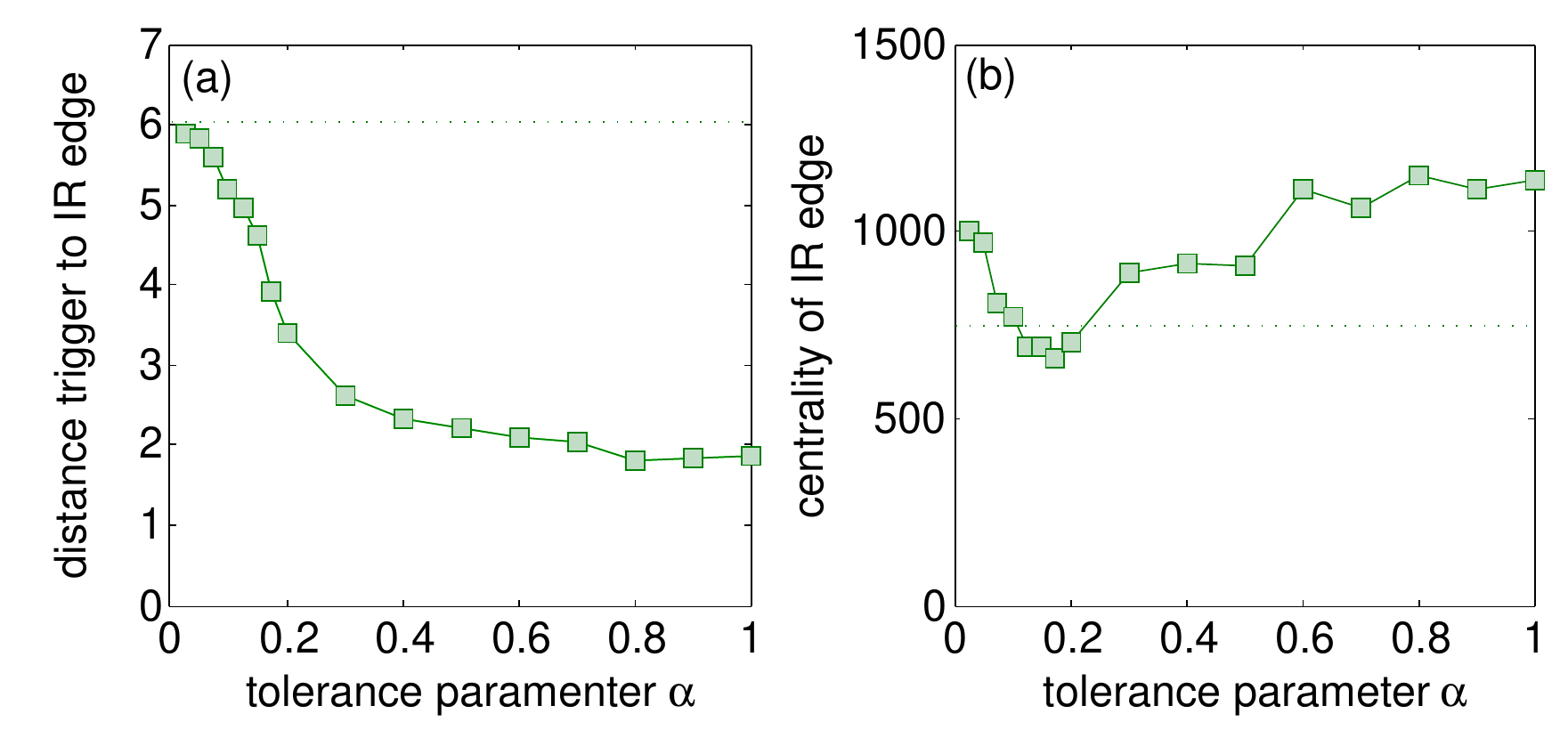}
\includegraphics[width=8cm, angle=0]{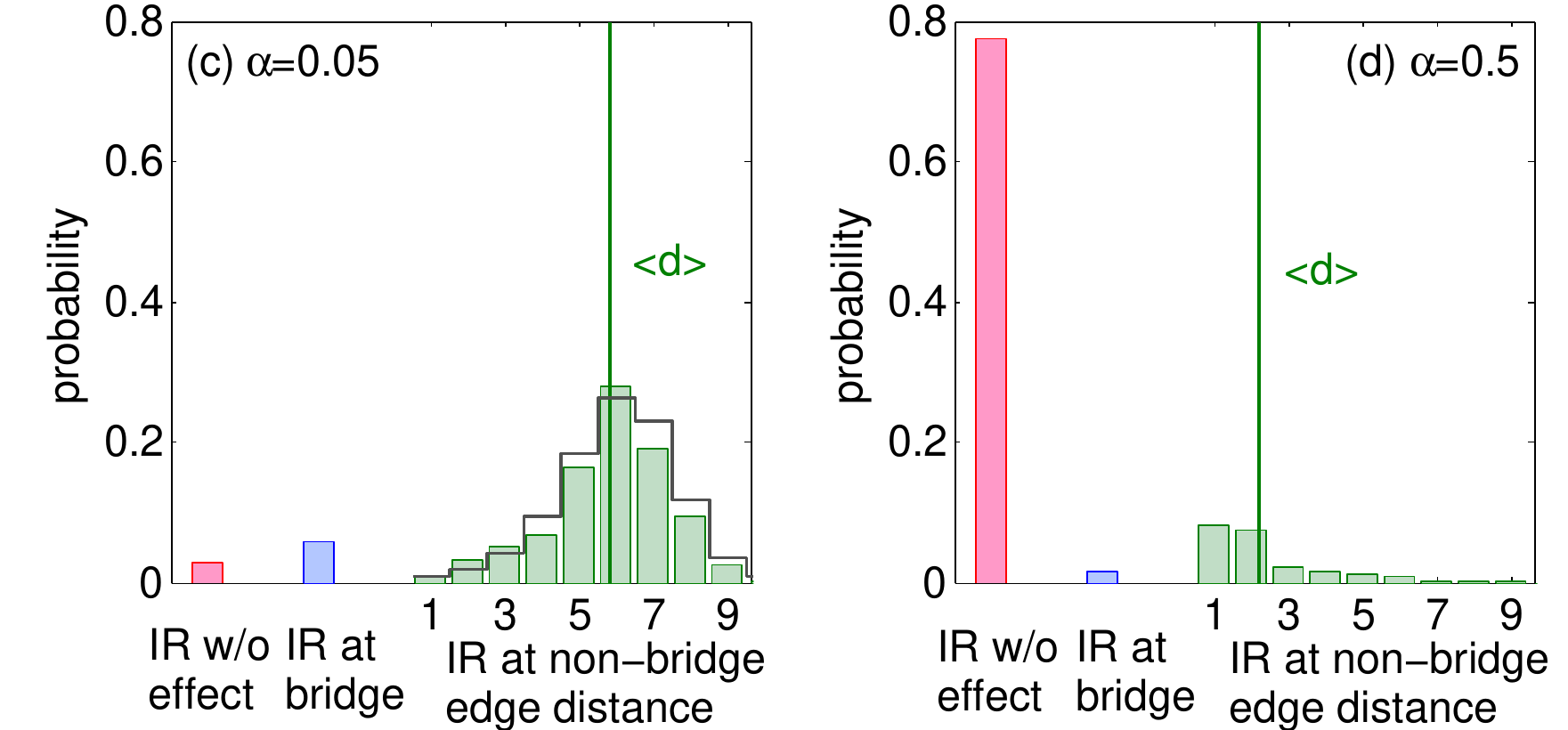}
\caption{\label{fig:prevent2}
(Color online)
Nonlocality of intentional removal for preventing cascading failures.
(a) Average distance of the intentionally removed edge and the
trigger edge as a function of the tolerance parameter 
$\alpha$.
(b) Average betweenness centrality of the intentionally removed edge
as a function of the tolerance parameter $\alpha$.
In (a,b) we disregard cases where IR has no effect 
or disconnects the grid. The dashed line shows 
the average shortest path distance $L$ and the average
centrality, respectively, for comparison.  
(c,d) Histogram of the distances of the intentionally removed edge
and the trigger edge in case that IR leads to an increase of the 
final cluster size without disconnecting the network (green bars)
for two values of $\alpha$
Grey lines show the distance distribution for all edges in the
network for comparison.
The red bars indicate the probability that that IR has no 
effect and the blue bars the probability that IR disconnects 
the grid.
Results are averaged over all trigger edges for a
W/S network with $N=500$, $k=4$ and
$q=0.2$. The dashed lines  
}
\end{figure}

There is a further significant difference between networks with high and
low redundancy, respectively.
In Fig.~\ref{fig:prevent2} we analyze the characteristics of the 
intentionally removed edge which optimizes $G/G_0$. 
The betweenness centrality of the intentionally removed 
edges is higher than average, except for intermediate values of the 
parameter $\alpha$ (cf.~\cite{Mott04}).
Similar results are found for the closeness centrality (not shown).
Most interestingly, the distance of the intentionally removed edge to the respective 
trigger edge decreases significantly with $\alpha$. 
In the case of low $\alpha$ the distance is approximately equal to
the average shortest path distance $L$, but for high $\alpha$ the
distance is much smaller. In this case, cascades propagate mostly 
locally such that they can be stopped by \emph{local} countermeasures.

This finding is further explicated in Fig.~\ref{fig:prevent2} (c,d) where
we plot a histogram of the distance removed-to-trigger 
as well as the probability that IR has no effect for two values of 
the tolerance parameter $\alpha$. 
For $\alpha=0.05$, the distribution of the distance
removed-to-trigger closely resembles the distribution of the distance
of two arbitrary edges.
This observation imposes the conclusion that the location of the 
intentionally removed and the trigger edges are uncorrelated
to a large extend. 
On the contrary, the distribution of distances decreases monotonically
with a small average for $\alpha=0.5$.

\section{Conclusion}

Large-scale outages in complex supply networks are often caused 
by cascades of failures triggered by the breakdown of a single 
element of the network. It is thus essential to understand 
the propagation of cascades in order to improve the stability 
of the power grids and the security of our electric power supply. 

In this article we have analyzed cascading failures in an elementary
topological model introduced by Motter and Lai \cite{Mott02}
from a microscopic perspective. 
We have shown that \emph{nonlocal failures} occur regularly
for general network topologies within this model. Such events are hard 
to predict theoretically and potentially hard to prevent in practice. 
Remarkably, nonlocal effects are strongly suppressed in networks 
with a high clustering and small average path length. 
In such networks, including many examples from power grids to 
biological and social networks, cascades propagate predominantly locally, 
i.e.~from one edge to an adjacent one.

One particularly effective countermeasure to stop or contain 
cascades is the intentional removal (IR) of a carefully selected 
additional edge \cite{Mott04}. Two very different microscopic 
scenarios were found depending on the tolerance parameter $\alpha$, 
which measures the global redundancy of the grid.
If the tolerance parameter $\alpha$ is small such that the network is 
vulnerable to cascades, IR must be applied on a global scale. That is, 
the optimum edge to be removed is generally located at a large 
distance to the initially failing edge.
On the contrary, cascades propagate mostly locally in highly
redundant networks (large $\alpha$) such that local countermeasures 
are  generally sufficient.

\begin{acknowledgments} 

We gratefully acknowledge support from  the Helmholtz Association 
(grant no.~VH-NG-1025 to D.W.), 
the Federal Ministry of Education and Research (BMBF grant nos.~03SF0472B
and 03SF0472E to M.T. and D.W.)
and the Max Planck Society to M.T.

\end{acknowledgments} 



\begin{thebibliography}{30}
\bibitem{footnote}
A triangle is defined as a set of three vertices 
$(a,b,c)$, which are mutually completely connected. That
is, there are \emph{three} edges $(a,b)$, $(a,c)$ and $(b,c)$
connecting the edges.
A connected triplet is defined as a set of three vertices 
$(a,b,c)$, which are mutually connected by exactly \emph{two} 
edges. A connected triplet can be seen as a triangle with one
missing edge.
\end{thebibliography}

\end{document}